\documentclass[prl,showpacs,aps,twocolumn]{revtex4}
\usepackage{amsmath}
\usepackage{graphicx}
\usepackage{dcolumn}
\usepackage{amssymb}
\usepackage{color}

\usepackage{graphicx,comment}
\usepackage{epstopdf}
\usepackage{bm}
\usepackage{enumerate}
\usepackage{fancyhdr,color}
\usepackage{verbatim}
\usepackage{amsmath,amsthm}
\usepackage{amsfonts}
\usepackage{hyperref}

\def\mg{{\mathfrak g}}
\def\hh{{\hat h}}
\def\one{{\mathchoice {\rm 1\mskip-4mu l} {\rm 1\mskip-4mu l} {\rm
1\mskip-4.5mu l} {\rm 1\mskip-5mu l}}}
\newcommand{\ket}[1]{\left| #1\right\rangle}        
\newcommand{\sket}[1]{| #1\rangle}  
\newcommand{\sbra}[1]{\langle #1|}  

\begin{document}


\title{A Trotter-Suzuki approximation for  Lie groups with applications to Hamiltonian simulation}

\author{
Rolando D. Somma}
\affiliation{Theoretical Division, Los Alamos National Laboratory, Los Alamos, NM 87545, US.}

\date{\today}

  \begin{abstract}
  We present a product formula to approximate the exponential of a skew-Hermitian operator that is a sum
  of generators of a Lie algebra. The number of terms in the product depends on the structure
  factors. When the generators have  large norm with respect to the dimension of the Lie algebra, 
  or when the norm of the effective operator resulting from nested commutators is less than the product of the norms,
  the number of terms in the product is significantly less than that obtained from well-known results.
  We apply our results to construct product formulas useful for the quantum simulation
  of some continuous-variable and bosonic physical systems, including systems whose potential
  is not quadratic. For many of these systems, 
  we show that the number of terms in the product  can be
  sublinear or even subpolynomial in the dimension of the relevant local Hilbert spaces, 
  where such a dimension
  is usually determined by the energy scale of the problem. Our results 
  emphasize the power of quantum computers for the simulation of 
  various quantum systems.
  \end{abstract}
  
  
  \maketitle
  

The simulation of quantum systems is one of the most promising applications of quantum computers~\cite{Fey82,Llo96}.
A main challenge to devise quantum algorithms for physics simulation 
is to approximate the evolution operator, $U:=e^{-iHt}$, as a sequence of simple gates. Typically,
this is done by using a product formula based on the so-called Trotter-Suzuki
approximation~\cite{Suz90,Suz91}. When the Hamiltonian $H$ is expressed as a sum of ``simple'' Hamiltonians,
such a product can be interpreted as an approximation of $U$
by short-time evolutions under each simple Hamiltonian. The complexity of the algorithm
is then related to the number of terms in the product. When the Hamiltonians are bounded and 
time independent, this number is $O(\exp(1/\eta) (\| H \| t)^{1+\eta})$ in the best case, 
for arbitrary small $\eta >0$~\cite{BAC07,WBH+10}.
More recently, a new method for simulating the evolution operator
was introduced in~\cite{BCC+14,BCC+15}. This method approximates $U$ by 
implementing a truncated Taylor series of the exponential.
When the Hamiltonians are bounded and under some additional assumptions,
the complexity of this method is $\tilde O(\| H \| t)$. (The $\tilde O$ notation
hides logarithmic factors.) This complexity is almost linear in the evolution time
and can be shown to be optimal, i.e., there is a matching lower bound~\cite{BCC+14}.
Methods for approximating $U$ are also useful for, e.g., simulating physical systems with Monte-Carlo and other
classical methods,
and for simulating differential equations with the split-step Fourier method~\cite{TA84}.


Several works study the potential of the above methods in particular examples,
such as quantum chemistry and physical systems with various particle statistics~(c.f., \cite{OGKL01,SOGKL02,ADL+05,KWP+11,WBCH14,PHW+14,BBK+15,BBK++15}).
However, a main inconvenience of the methods 
in~\cite{BAC07,WBH+10,BCC+14,BCC+15} and other related methods (c.f.,~\cite{AT03,BC12,CW12}) is that they cannot be directly
applied to the case of, for example, unbounded operators, or can lead to unnecessary
complexity overheads. These methods also consider 
the worst case scenario and do not exploit certain structures of the problem, such
as commutation relations between the simple Hamiltonians. For example,
consider the case $H=\hat J_x + \hat J_y$, where $\hat J_\alpha$, $\alpha=x,y,z$,  are the well-known
$\mathfrak {su}(2)$ angular momentum operators acting on a (spin) system of dimension $2J+1$.
In this case, $\|\hat J_{\alpha}\| = O(J)$ and the  results in~\cite{Suz90,Suz91,BAC07} would
yield an approximation of $U$ as a product of a polynomially large (in $J$)
number of exponentials of $J_x$ and $J_y$.
However, one can exactly decompose $U$ (up to a phase) using three exponentials
by means of Euler-angle decompositions~\cite{Dal08} or obtain 
a very good approximation of $U$ with a number of exponentials that is subpolynomial
in $J$, which is a consequence of our main results described below.

In this paper, we build upon the results in~\cite{Suz90,Suz91,BAC07}
and consider the case in which the simple Hamiltonian terms belong to a certain Lie algebra.
By exploiting the structure in commutation relations, our main result is a significantly improved
bound on the number of terms in the product formula that approximates $U$. 
We illustrate our main result with several examples. The first example regards the quantum harmonic oscillator (QHO),
where the operators in the Hamiltonian generate a Lie algebra of dimension 3. For the QHO,
we show that $U$ can be approximated by a sequence of simple unitaries of length subpolynomial in the dimension 
of the relevant Hilbert subspace, recovering a result in~\cite{Som15}. The second example
regards  coupled QHOs and we show that the number of exponentials 
in the approximation of $U$ is also subpolynomial in the dimension of the local Hilbert subspaces. 
The third example  regards a one-dimensional quantum system where, unlike the QHO, the potential is not necessarily
quadratic. Depending on the form of the potential, the number of terms in the approximation
of $U$ can be sublinear or subquadratic in the dimension of the relevant Hilbert subspace.

A common feature in all these examples is that the norm of the effective operator
resulting from nested commutators of operators in the Lie algebra 
can be shown to be significantly smaller than the product of 
the norms of all effective operators appearing in such commutators. The effective operator is basically the operator
 projected on a relevant and finite dimensional Hilbert subspace. In our examples, the dimension of such subspaces
is typically determined by an energy scale associated with the problem. While we do not construct quantum algorithms 
for simulating $U$, our results suggest that quantum computers can simulate
 the evolution of several continuous-variable quantum systems
more efficiently than conventional computers. A step in this direction was recently given in~\cite{Som15}, where we provided
a quantum algorithm for simulating the QHO with subpolynomial complexity.
Classical algorithms
for these problems are expected to have a worst-case complexity that 
is polynomial (e.g., worse than quadratic) in the dimension of the Hilbert subspaces,
as one has to deal with matrices of polynomial dimension.

{\em Additional related work.--}
A detailed analysis of the approximation
error induced by the so-called second order Trotter-Suzuki approximation, in terms of commutators,
was recently done in~\cite{PHW+14}
for the quantum chemistry problem, and subsequently analyzed in~\cite{BMW15}.
The resulting number of terms in the approximation
is still scales with $\|H\|$ for that case. In contrast,
our work is more concerned with problems where the 
the norm of the effective operators can be large, as in the case
of the quantum simulation of continuous-variable quantum systems.
Our goal is to provide a product formula where the number of terms
can be sublinear in the norm of the effective Hamiltonian.

\vspace{0.1cm}

We define the problem and state our main results in more detail. 
Some applications for the simulation of relevant quantum systems are discussed later.

\vspace{0.2cm}

{\em Problem statement.--}
We let $\mg$ be a real Lie algebra of infinite or finite dimension $K$ with basis $\{ \hat h_1,\ldots,\hat h_K\}$. 
Since we are interested in the case where $U$ is unitary, we assume that $\hh_k$ is skew-Hermitian for all $k$.
The Lie bracket is  $[\hh_k,\hh_{k'}]:=\hh_k \hh_{k'} - \hh_{k'} \hh_k$ and
\begin{align}
\label{eq:structure}
[\hh_k,\hh_{k'}] = \sum_{k''=0}^K \gamma^{k,k'}_{k''} \hh_{k''} \;.
\end{align}
The constants $\gamma^{k,k'}_{k''} \in \mathbb R$ are the structure factors of $\mathfrak g$.
We let $X:= \sum_{k=1}^L  \hh_k$, where $L \le K$ with no loss of generality and $L<\infty$. 
Given a precision parameter $\epsilon >0$,  evolution time $t \ge 0$, and initial state $\ket \psi$, the goal is to approximate $U:=e^{t X}$ by
a unitary  $W$ such that
\begin{align}
\label{eq:firstapprox}
\| (e^{ t X} - W) \ket \psi \| \le \epsilon \;.
\end{align}
$\| \ket \phi \|$ is the Euclidean norm of the state $\ket \phi$. $W$ admits the decomposition
\begin{align}
\label{eq:productform}
W=\prod_{n=1}^N e^{t_n \hh_{k_n} } \;,
\end{align}
where $t_n \in \mathbb R$ and $k_n  \in [L]:=\{1,2, \ldots, L\}$.

In the following, the maximum is always 
taken over $\lambda,\lambda' \in [0,t]$ and $k_i \in [L]$ unless  noted explicitly.

\vspace{0.2cm}

{\em Main results.--} 
Let   $r \ge 1$ be an integer such that
\begin{align}
\label{eq:firstresult}
2 (N_p)^2  \sum_{j=2p}^\infty  (f_j N_p/r)^{j} t^{j+1} \beta_{j+1 } \le \epsilon \;,
\end{align}
where $N_p:=2L5^{p-1}$, $p \ge1$ is an arbitrary integer,  $\beta_{j} := \max 
 \| [\hh_{k_1},[\ldots,\hh_{k_{j}}] \ldots ] U(\lambda) \ket \psi \| $, and $f_j:=2$ if $j<N_p$ or $f_j: = 6N_p/j$ if $j \ge N_p$.
Then, there is a unitary $W$ that approximates $U$ as in Eqs.~\eqref{eq:firstapprox} and~\eqref{eq:productform}
 and the number of terms in the product   is $N= r N_p$.

When the dimension of $\mathfrak g$ is finite and if $yt \ge \epsilon$, then
\begin{align}
\label{eq:secondresult}
N =O \left ( { \; 5^{2p} L^{2+\frac 1 p} \; \beta \; y^{\frac 1{2p}} \; t^{1+\frac 1{2p}}}/{\epsilon^{\frac 1{2p}}} \right ) \;,
\end{align}
where
\begin{align}
\label{eq:betadef}
\beta := \max_{k,k' \in \mathfrak g} \sum_{k''=1}^K |\gamma^{k,k'}_{k''}| 
\end{align}
 and $y:= \max \| \hh_k U(\lambda)\ket \psi \|$. 


\vspace{0.2cm}
{\em Proofs.--}
Following Suzuki~\cite{Suz90,Suz91}, we define the unitary
\begin{align}
\nonumber
W_2(\lambda) = \prod_{k=1}^L e^{\lambda \hh_k /2}  \prod_{k=L}^1 e^{\lambda \hh_k  /2} \;,
\end{align}
and the recursion relation (for integer $p\ge1$)
\begin{align}
\nonumber
W_{2p+2}(\lambda) =(W_{2p}(s_{p} \lambda))^2 W_{2p}((1-4s_{p})\lambda) (W_{2p}(s_{p} \lambda))^2 \;.
\end{align}
The constants are $s_p = 1/( {4 - 4^{1/(2p+1)}})$. Similarly, we can write $W_{2p}(\lambda) = V_{N_p} \ldots V_1$, where 
each unitary $V_n$ is of the form $e^{(s'_n \lambda) \hh_{k_n}}$ and ${k_n} \in [L]$. The number of unitaries in the product
results from the recursion and is $N_p = 2L5^{p-1}$, and the coefficients $s'_n \in \mathbb R$ satisfy $|s'_n| < 1$.

The operator $\epsilon_{2p}(\lambda): = W^\dagger_{2p}(\lambda) U (\lambda)- \one$, where $U(\lambda) := e^{\lambda X}$ and $\lambda \in [0,t]$,
will provide information about the accuracy of the approximations $W_{2p}$. 
Our first goal is then to find an upper bound of $\varepsilon_{2p}(\lambda):=\| \epsilon_{2p}(\lambda) \ket \psi \|$, where $\ket \psi$ denotes some initial
quantum state.
Since $\epsilon_{2p} (\lambda)= \int_0^\lambda d \lambda' \; \partial_{\lambda'} \epsilon_{2p}(\lambda')$, we obtain
$\epsilon_{2p} (\lambda)
 =\int_0^\lambda d \lambda' \; W^\dagger_{2p}(\lambda')  f_{2p}(\lambda') U(\lambda') $
and then 
\begin{align}
\label{eq:eps2p}
\varepsilon_{2p}(\lambda) \le \lambda \max \| f_{2p}(\lambda') U(\lambda') \ket \psi \| \;.
\end{align}
The operator $f_{2p}(\lambda)$ can be obtained from the chain rule:
\begin{align}
\label{eq:f2pA}
f_{2p}(\lambda)=X- \sum_{n=1}^{N_p}s'_n V_{N_p}\ldots V_{n+1} \hh_{k_n} V_{n+1}^\dagger \ldots V_{N_p}^\dagger \;.
\end{align}
From the Lie algebra property, $V_n \hh_k V_n^\dagger = \one + (s'_n \lambda)[\hh_{k_n},\hh_k]+ (s'_n \lambda)^2[\hh_{k_n},[\hh_{k_n},\hh_k]]/2 + \ldots$.
Then,  Eq.~\eqref{eq:f2pA} is a combination of nested commutators
of those $\hh_k$ appearing in $X$ so that
\begin{align}
\label{eq:f2p}
f_{2p}(\lambda') = \sum_{k=1}^K c_k(\lambda') \hh_k \;.
\end{align}
The results in~\cite{Suz90} imply that the lowest degree of the Taylor series of $\epsilon_{2p}(\lambda)$, for $\lambda \rightarrow 0$,
is $2p+1$. It follows that the   lowest degree in a Taylor series of the coefficients $c_k(\lambda') \in \mathbb R$ is $2p$ and 
\begin{align}
\label{eq:f2p2}
f_{2p}(\lambda') = \sum_{j=2p}^\infty \lambda'^j \hat r_j \;.
\end{align}
Each $\hat r_j \in \mg$  results from sums of nested commutators of length $j+1$, e.g., $[\hh_{k_1},[\hh_{k_2},[\ldots,\hh_{k_{j+1}}]\ldots]]$,
with each $k_i \in [L]$.
The maximum number of possible nested commutators of such length  involved in $\hat r_j$ is bounded by
\begin{align}
\label{eq:partition}
N_p \begin{pmatrix}
N_p + j  \cr j +1
\end{pmatrix} \;.
\end{align}
The factor $N_p$ results from the sum of at most $N_p$ transformations of the $\hh_k$ when using Eq.~\eqref{eq:f2pA}
 and the binomial coefficient is the number of possible
ways of partitioning $j+1$ in $N_p$ parts.

Since $N_p \ge 5$ for $p\ge 2$, and $j \ge 2$, Eq.~\eqref{eq:partition} can be  
bounded by $2^{j+1} (N_p)^{j+2}$. 
 When $j+1 \le N_p$, we can then obtain 
$\| \hat r_j U(\lambda ') \ket \psi \|  \le 2^{j+1}( N_p )^{j+2}  \beta_{j+1}$,
where 
\begin{align}
\nonumber
\beta_{j} = \max
 \| [\hh_{k_1},[\ldots,\hh_{k_{j}}] \ldots ] U(\lambda') \ket \psi \| 
\end{align}
is strongly dependent on the structure of the algebra. When $j+1 >N_p$, we can obtain an improved bound 
because nested commutators of length $j+1$ must involve the same $\hh_{k}$ more than once.
Let $j+1 = lN_p +l'$, with $l \ge 1$ and $l'<N_p$ being nonnegative integers. Then, in determining each $\hat r_j$ there is also a constant factor
that is bounded  by $(1/(l+1)!)^{l'} (1/l!)^{N_p-l'}$ due to the Taylor series  of each transformation $V_{N_p}\ldots V_{n+1} \hh_{k_n} V_{n+1}^\dagger \ldots
V_{N_p}^\dagger$ in Eq.~\eqref{eq:f2pA}.
This constant corresponds to the case in which the nested commutator of length $j+1$ results from the $(l+1)$th order in the Taylor series 
 of $l'$ operators $V_n$ and the $l$th order in the Taylor series  of the remaining $N_p-l'$ operators $V_n$.
%
This factor is easily bounded by $(1/l!)^{N_p}$
and then $\| \hat r_j U(\lambda ') \ket \psi \|  \le (1/l!)^{N_p} 2^{j+1}( N_p )^{j+2}  \beta_{j+1}$. We can use Stirling's approximation
and $l \le j/N_p$
to obtain $1/l! \le (e/l)^l \le (3 N_p/j)^{j/N_p}$.
These bounds together with Eqs.~\eqref{eq:eps2p} and~\eqref{eq:f2p2} now imply
\begin{align}
\label{eq:errorbound1}
 \varepsilon_{2p}(\lambda)    \le 2 (N_p)^2\sum_{j=2p}^\infty  \lambda^{j+1} (f_j N_p)^{j} \beta_{j+1 } \;,
\end{align}
where $f_j=2$ if $2p \le j < N_p$ and $f_j=6 N_p/j$ if $j \ge N_p$. 
%
%
%
%
The case of $\beta_j=0$, for all $j$, corresponds to a commutative algebra and the error is exactly 0 in that case.
The interesting case is when some $\beta_j>0$ and from now on we assume that  there exists $\lambda>0$ such that
Eq.~\eqref{eq:errorbound1} converges and is bounded.

To find an approximation of $U=e^{tX}$, $t 
\ge 0$, we split $t$ into $r$ segments of 
size $\lambda = t/r$.
The subadditivity property of errors implies $\| (U-W) \ket \psi \| \le r \varepsilon_{2p}(t/r)$,
where we defined $W:=(W_{2p}(t/r))^r$.
Then, $r \varepsilon_{2p}(t/r) = 2 (N_p)^2  \sum_{j=2p}^\infty  (f_j N_p/r)^{j} t^{j+1} \beta_{j+1 }$ and
for precision $\epsilon>0$, it suffices to satisfy $r \varepsilon_{2p}(t/r) \le \epsilon$ [Eq.~\eqref{eq:firstresult}].
The total number of exponentials in $W$ is $N=r N_p$;
this proves our first result.

When the dimension of the Lie algebra is finite, it is useful to obtain
$\beta$ as in Eq.~\eqref{eq:betadef} and 
$y= \max \| \hh_k U(\lambda')\ket \psi \|$.
It follows that $\beta_{j+1} \le \beta^j y$ and, since $f_j \le 6$ for all $j$, Eq.~\eqref{eq:errorbound1} implies
\begin{align}
\varepsilon_{2p}(\lambda) \le (N_p y/\beta) \sum_{j=2p}^{\infty} (6 \lambda N_p \beta)^{j+1} \;.
\end{align}
To satisfy $r \varepsilon_{2p}(\lambda) \le \epsilon$, it suffices to choose
\begin{align}
\label{eq:Rfinite}
r =\left \lceil  { \; 5^{p+2} \; L^{1+ \frac 1 p} \; \beta \; y^{\frac 1{2p}} \; t^{1+\frac 1{2p}}}/{\epsilon^{\frac 1{2p}}} \right \rceil \;.
\end{align}
 This assumes that $yt \ge \epsilon$ so that $\lambda=t/r$ is sufficiently small for Eq.~\eqref{eq:errorbound1}
to converge.
%
Multiplying Eq.~\eqref{eq:Rfinite} by $N_p$ gives $N$ in Eq.~\eqref{eq:secondresult} and proves our second result.

Below we obtain $N$  for the approximation of the evolution operator of various quantum systems.
 
 \vspace{0.2cm}
 
{\em Applications.--}
Similar results to those in \cite{BAC07} can be essentially recovered if we assume that each $\hh_k$ in $X$ is a bounded operator
acting on a finite-dimensional Hilbert space \footnote{The number of exponentials given by our Eq.~\eqref{eq:secondresult}
can actually be smaller than that obtained in \cite{BAC07}. For example, when $\|\hh_k \|\le 1$ and $\epsilon$ and $t$
are constant,
we obtain $N=O(5^{2p} L^{2+1/p})$ while \cite{BAC07} implies $N=O(5^{2p} L^{3+1/2p} )$.}. 
Those results consider the worst-case scenario and do not exploit
certain structures of the commutation relations in the algebra.
Thus, to emphasize the importance of our results,
we first  provide a Trotter-Suzuki approximation for certain finite-dimensional 
Lie algebras that is well suited to the case of continuous-variable quantum systems.
In all our examples, $X:=-iH$, where $H$ is the Hamiltonian of the system. Then $U$
corresponds to the evolution operator and our goal is to find a product formula that approximates it.

We consider first the QHO, $H=(\hat p^2+ \hat x^2)/2$, where $\hat p$ and $\hat x$ are the momentum
and position operators, respectively ($\hbar =1$). 
The operators $i \hat p^2$ and $i \hat x^2$, together with $i\{\hat x,\hat p\}:=i\hat x \hat p+i\hat p \hat x$, 
are a basis of the $\mathfrak {sp}(2)$ Lie algebra of dimension $K=3$: $[i\hat x^2, i\hat p^2]=-2i\{ \hat x,\hat p\}$, $[i\hat x^2,i\{\hat x,\hat p \}]=-4i \hat x^2$, and 
$[i\hat p^2 ,i \{\hat x, \hat p\}]=4i \hat p^2$.  These commutation relations easily follow from the canonical
commutation relation $[\hat x,\hat p]=i$, and then $\beta=O(1)$.
The results in \cite{BAC07} cannot be directly applied to this case as $\hat x$ and $\hat p$ are unbounded operators.
If $\ket \psi=\sum_{m=0}^{m'} c_m \sket{\phi_m}$, where $\sket{\phi_m}$ are normalized eigenstates of $H$
of eigenvalue $m+1/2$, then $y \le \max_{k,m} \| h_k \sket {\phi_m} \|$ and $y=O( m')$. This follows from the well-known
properties of $\hat x$ and $\hat p$, where $\hat x \sket{\phi_m}=( \sqrt m \sket{\phi_{m-1}} + \sqrt{m+1} \sket{\phi_{m+1}})/\sqrt 2$ and
$\hat p \sket{\phi_m}=-i ( \sqrt m \sket{\phi_{m-1}} - \sqrt{m+1} \sket{\phi_{m+1}})/\sqrt 2$. 
For precision $\epsilon$, the number of terms in the approximation $W$ of $U$ results from Eq.~\eqref{eq:secondresult} and 
simple calculations show
\begin{align}
\label{eq:QHO}
N   =O \left( 5^{2p} \;  (m'/\epsilon)^{\frac 1{2p}} \; t^{1+ \frac 1{2p}}  \right ) \;,
\end{align}
where we also used $L=O(1)$.
We can choose an optimal value of $p$ that minimizes the value of $N$ in Eq.~\eqref{eq:QHO}.
This occurs when $p \approx \sqrt{\log(m' t/\epsilon)/\log (5)}$ and then $N =O(t \exp(\sqrt{\log(m' t/\epsilon)}))$.
Note that $N$ is subpolynomial in $m'$, i.e., $N/(m')^\alpha$ approaches 0 in the limit of large $m'$ for any $\alpha>0$. This result suggests
that a subexponential quantum speedup can be attained in a quantum-computer simulation of $U$ in the gate model.
We showed that this is  possible in~\cite{Som15}.

We now consider the more general case of $M$ coupled QHOs, where the Hamiltonian is, for example,
\begin{align}
\nonumber
H = \frac 1 2 \sum_{l=1}^M (\hat p_l^2 + \hat x_l^2) - \sum_{l \ne l'} \hat x_l \hat x_{l'} \;.
\end{align}
It is well known that the operators appearing in $X$ generate the $\mathfrak {sp}{(2M)}$ Lie algebra
of dimension  $K=M(2M+1)$. The structure factors of the algebra follow from the canonical commutation
relations $[\hat x_l,\hat p_{l'}]=i \delta_{ll'}$ and $[\hat x_l,\hat x_{l'}]=[\hat p_l,\hat p_{l'}]=0$, where $\delta_{ll'}$
is the Kronecker delta. As in the previous case, these factors imply $\beta =O(1)$~\cite{RN92}.
 With no loss of generality, the evolved state $U(\lambda) \ket \psi$ is a linear combination of
states $\sket{\phi_{m_1},\ldots,\phi_{m_M}}$. We will assume that there is $m'$ such that, if we set $m_i \le m'$ for all $i$ and $\lambda$, then
the approximation error induced by this assumption in the evolved state is negligible. Note that $m'$
determines a local  ``energy scale'', as the expected value of $(\hat p_l^2 + \hat x_l^2)/2$ in the evolved state is $O(m')$.
Then $y = O(m')$, $L=O(M^2)$,  and the number of terms in the approximation $W$ of $U$ given by Eq.~\eqref{eq:secondresult} is
\begin{align}
\nonumber
N   =O \left(  { 5^{2p} \;  (m'/\epsilon)^{\frac 1{2p}} \; (M^4t)^{1+ \frac 1{2p}}}  \right ) \;.
\end{align}
The optimal value of $p$ that minimizes $N$ can be obtained as before
and gives $N = O(M^4 t \exp(\sqrt{\log(m' Mt/\epsilon)}))$, which is subpolynomial in $m'$.
We note that a similar analysis and result applies for those $X$ (and corresponding Hamiltonians)
that are more general linear combinations of the generators of $\mathfrak {sp}{(2M)}$.

To demonstrate our result when the dimension of the Lie algebra is infinite, we apply our 
bound to the case where $H = (\hat p^2  + \hat x^q)/2$, for integer $q > 2$.
The Lie algebra generated by $i\hat p^2$ and $i\hat x^q$ is infinite dimensional.
To bound the errors in the approximation of $U$, it is necessary to study the properties of nested commutators
for this case.
We note that $[\hat x^k, \hat x^l \hat p^m]$ are polynomials of $\hat x$ and $\hat p$.
We will use induction to show that the degree of this polynomial
is $k+l+m-2$ and, in particular, the degree associated with $\hat x$ is $k+l-1$
and the degree associated with $\hat p$ is $m-1$.
When $k=m=1$, we have $[\hat x,\hat x^l \hat p] = \hat x^l [\hat x,\hat p]=i \hat x^l$, so the statement is valid in this case.
Also, it is simple to show $[\hat x^k,\hat x^l \hat p^{m+1}] = [\hat x^k,\hat x^l \hat p^{m}] \hat p + i k \hat x^l \hat p^m
\hat x^{k-1}$ and $[\hat x^{k+1},\hat x^l \hat p^{m}] = \hat x [\hat x^k,\hat x^l \hat p^{m}]  + i l \hat x^l \hat p^{m-1}
\hat x^{k}$, so that increasing $k$ or $m$ by 1 only increases the degree of $\hat x$ or $\hat p$ by 1, respectively.
This demonstrates the induction step.
These  properties also imply that if we commute a polynomial in $\hat x$ and $\hat p$ with $\hat p^2$,
then the degree of $\hat x$ is reduced by 1 while the degree of $\hat p$ is increased by 1.
Also, if we commute such a polynomial with $\hat x^q$, the degree of $\hat x$ is increased by $q-1$ while
the degree of $\hat p$ is reduced by 1. This is a useful observation because 
the operators $\hat r_j$ in Eq.~\eqref{eq:f2p2} result from nested commutators of $i\hat x^q$ and $i\hat p^2$ in this case.
In particular,
the only nonzero nested commutators of length $j+1$ are those for which the number
of appearances of $\hat x^q$ is less than or equal to $(j +2)/2$; otherwise the degree of $\hat p$
would be negative leading to an inconsistency. Then, the largest degree of the polynomial
generated by a nested commutator of length $j+1$ is upper bounded by $d_j=(q-2)(j+2)/2 +2$,
which was obtained for the worst case scenario in which the number of appearances of 
$\hat x^q$ is, at most, $(j+2)/2$. 

With no loss of generality, $U(\lambda')\ket \psi=\sum_m c_m(\lambda') \sket{\phi_m}$ and we 
assume that there exists $m'$ such that, if we cut off the sum at $m \le m'$, the error induced by 
this approximation is negligible or $O(\epsilon)$ for all $\lambda'\in[0,t]$. We will 
then bound the approximation error by assuming that $U(\lambda')\ket \psi$
has no support in those states $\sket{\phi_m}$ for $m>m'$. 
This poses no problem as the additional error factor in $\varepsilon_{2p}$
is still $O(\epsilon)$. The value of $m'$ determines the relevant energy scale of the problem-- see below.
 Then,
\begin{align}
\nonumber
\beta_{j+1} & \le 2^j \max_{ 0 \le l \le d_j, 0 \le m \le m' }
 \| \hat T_{l}  \sket{ \phi_m} \| \\
 \label{eq:betabound2}
 & \le 2^{3j/2} (m' + d_j)^{d_j/2}  \;,
\end{align}
where $T_{l}$ is a product of $l$ operators $\hat x$ and $d_j-l$ operators $\hat p$, in some order.
To bound the error in the approximation of $U$ by $W$ we first use
Eq.~\eqref{eq:betabound2} in Eq.~\eqref{eq:errorbound1}. This error depends on $q$ and, for large $q$, 
it can diverge. Nevertheless, if we assume $2 \le q \le 6$,
 a few algebraic manipulations using Eq.~\eqref{eq:errorbound1}
imply 
\begin{align}
\nonumber
\varepsilon_{2p}(\lambda) \le \sum_{j=2p}^\infty  (c \lambda N_p^2)^{j+1} (m')^{\frac{j(q-2)}4+ \frac q 2} \;,
\end{align}
where $c>1$ is a constant. We can obtain $r$ by setting $r \epsilon_{2p}(t/r) \le \epsilon$ and then use it to obtain $N$.
Then, the number
of exponentials in $W$ in this case is
\begin{align}
\label{eq:Nx^q}
N = O \left(  {5^{3p} (m')^{\frac q {4p} + \frac q 4 - \frac 1 2} t^{1+\frac1{2p}}  } / \epsilon^{\frac 1 {2p}}\right) \;.
\end{align}
We note that Eq.~\eqref{eq:Nx^q} is a much better bound for $N$ than that obtained
if we replace the exponent of $m'$ by $(q/2)(1+1/2p)$ in the same equation. Such an exponent
would be obtained if we assumed that the norm of the effective operator
 $\hat x^q$ could be replaced by $O(m'^{\frac q 2})$ and then use the results in~\cite{BAC07}.
Our result for $N$ also suggests a polynomial quantum speedup with respect to classical methods
for simulating the evolution operator, as the dependence of $N$ with $m'$ is sublinear 
when $q \le 5$ and subquadratic when $q=6$.

We note that both, the classical and quantum algorithm complexities, depend
on the same value of $m'$ that determines the dimension of the
relevant Hilbert subspace associated with the evolved state. 
 It may then be useful to understand the dependence of $m'$ in the special case
where $\ket \psi=\sket{\phi_m}$. For example, it is simple to show $(\hat x^2+\hat p^2)/2 \le \one + (\hat p^2 + \hat x^q)/2$ when $q$ is even.
Then, $\sbra{\phi_m} U^\dagger(\lambda) (\hat x^2+\hat p^2) U(\lambda) \sket{\phi_m} = O(m^{q/2})$ and we can use Markov's inequality
to show that the support of $U(\lambda) \sket{\phi_m}$ in the space spanned by $\sket{\phi_{m'}}$, for $m' =\Omega(m^{q/2})$, is bounded by an arbitrarily small constant.
(We can improve this bound by using inequalities that involve moments of higher order.)
We can use this value of $m'$ in Eq.~\eqref{eq:Nx^q} for this particular case.

%

\vspace{0.2cm}

{\em Conclusions.--} 
We presented an improved product formula to approximate
the evolution operator of various quantum systems
that exploits the structure of commutation relations of the associated Lie algebra.
We applied this formula to examples of bosonic quantum systems and the results
suggest that quantum-computer simulations of such systems can be done significantly more efficiently
than classically possible.

\vspace{0.2cm}

{\em Acknowledgements.--}
We thank C. Batista at LANL for enlightening discussions.
This work was performed under the auspices of the
U.S.  DOE  contract  No.  DE-AC52-06NA25396  through  the
LDRD program at LANL.


\end{document}